# Valley pair qubits in double quantum dots of gapped graphene

G. Y. Wu[1,2*], N.-Y. Lue[1], and L. Chang[1]

[1]Department of Physics, National Tsing-Hua University, Hsin-Chu 30013, Taiwan, ROC; [2]Department of Electrical Engineering, National Tsing-Hua University, Hsin-Chu 30013, Taiwan, ROC; [*]E-mail: yswu@ee.nthu.edu.tw.

**Abstract**

The rise of graphene opens a new door to qubit implementation, as discussed in the recent proposal of valley pair qubits in double quantum dots of gapped graphene (Wu et al., arXiv: 1104.0443 [cond-mat.mes-hall]). The work here presents the comprehensive theory underlying the proposal. It discusses the interaction of electrons with external magnetic and electric fields in such structures. Specifically, it examines a strong, unique mechanism, i.e., the analogue of the $1^{st}$-order relativistic effect in gapped graphene. This mechanism is state mixing free and allows, together with the electrically tunable exchange coupling, a fast, all-electric manipulation of qubits via electric gates, in the time scale of ns. The work also looks into the issue of fault tolerance in a typical case, yielding at $10^{o}K$ a long qubit coherence time (~O(ms)).

PACS numbers: 73.63.Kv, 71.70.Ej, 68.65.Pq, 76.20.+q

## I. Introduction

At the base of quantum computing[1] realization is the physical implementation of qubits – two-state quantum information units. The rise of graphene[2,3] has expanded the spectrum of materials suited to the implementation. Graphene is a two-dimensional material with the unique electron dispersion simulating that of Dirac particles. Because the dispersion has two degenerate, independent energy valleys, a novel degree of freedom (d.o.f.) - the valley state of an electron appears as a potential information carrier.[4] This opens a new path with distinctive advantages to encode quantum information - using "valley-based" qubits in quantum dots (QDs), as discussed in the recent proposal of valley pair qubits.[5] The feasibility of valley qubits in graphene rings[6] or valley control in other allotropes of carbon, e.g., carbon nanotube (CNT) QDs with disorder,[7] have also been discussed recently which are based on different implementations and physical mechanisms.

Valley pair qubits utilize "valley singlet/triplet states" in double quantum dot (DQD) structures of gapped graphene to represent the logic 0/1 states, respectively. Notably, they are characterized by a) scalability and fault-tolerance, b) all-electric manipulation via electric gates, and c) long coherence time, all being rather useful assets in qubit implementation. The principles underlying the qubits overlap with those developed for spin qubits, but with important differences.

In the spin case, for instance, the quantum dot (QD) approach (-using confined electron spins)[8] usually serves as the basis. Because of energy quantization, electron state coherence in QDs is generally enhanced. In addition, the exchange coupling between electrons in adjacent

QDs is electrically tunable, aiding greatly the realization of all-electric qubit/qugate manipulation. Upon the QD approach, various paradigm tactics may additionally be applied, including: utilization of spin pair qubits to attain scalability and fault-tolerance,[9-13] the Rashba mechanism of spin-orbit interaction (SOI) to achieve coherent, electric manipulation of a single spin,[14-16] and materials with weak SOI and vanishing hyperfine field (HF), such as graphene[17] or CNT[18], in order to resolve the problem of state relaxation/decoherence.

The QD approach and the aforementioned tactics altogether constitute a good framework for the development of analogous qubits. However, being solutions to separate issues, the various tactics are sometimes at odds with one another in a material-dependent way. For instance, where a strong Rashba mechanism is available, the HF or the SOI inevitably cause state relaxation/decoherence,[19-23] whereas in materials with weak SOI, qubit manipulation based on the mechanism of electron spin resonance is often employed or suggested[24,25] which involves an AC magnetic field. For this reason, varied materials, e.g., GaAs,[13,25,26] CNT[18], or InAs,[27] have been exploited, each suited to certain tactics, in the recent demonstration of spin qubits.

Interestingly, in the case of valley pair qubits in DQD structures of gapped graphene, these tactics can all be realized (in their "valley" version) without contradicting one another, thus providing a fascinating prospect of valley-based quantum computing.[5] The qubit manipulation here is based on a unique mechanism, namely, the analogue of the 1st-order relativistic effect due to the Dirac spectrum in gapped graphene. Being of "relativistic" origin, the mechanism is similar to the SOI, with strength comparable to that of the SOI in InAs. But there is a remarkable difference – it is valley-diagonal and hence state-mixing free, as opposed to the SOI. Based on the mechanism, a DC or AC electric field can be applied to modulate the orbital magnetic moment of a confined electron, creating a magnetic moment gradient in the DQD structure. In the presence of a static, uniform magnetic field, it results in a coherent qubit state rotation (in the effective Bloch sphere). For manipulation of the qubit into an arbitrary state, another rotation independent of the foregoing one is required, and provided by exploiting the exchange coupling in the qubit. With these two rotations together, a fast, all-electric qubit manipulation may be performed by standard electric gate operation, in the time scale ~ O(ns). In a typical case, the valley relaxation time is estimated to be O(ms), sufficiently long for the manipulation.

The work here presents the comprehensive theory underlying these valley pair qubits. The tuning of magnetic moment is one focus of this presentation. It is discussed within a Schrodinger-type equation including the $1^{st}$-order "relativistic correction" (R.C.). In addition, the work examines the issue of qubit coherence and investigates specifically the phonon-mediated valley singlet-triplet transition. The presentation is organized as follows. **Sec. II** presents an overview of the principles underlying valley pair qubits. **Secs. III~VI** supply the details. In particular, **Sec. III** provides the quantum-mechanical description of graphene QDs, including the Schrodinger-type equation with the "$1^{st}$-order R.C.". **Sec. IV** estimates the exchange interaction between two separately confined electrons in the DQD structure. **Sec. V** discusses the valley splitting in a QD, in the presence of a normal uniform magnetic field, and also derives the associated valley magnetic moment ($\mu_v$) correct to the "$1^{st}$-order R.C.". In **Sec. VI**, $\mu_v$-tuning in DC / AC modes is discussed in relation to the qubit manipulation. **Sec. VII** investigates the phonon-mediated valley relaxation and the valley singlet-triplet transition. **Sec. VIII** summarizes the study. **Appendix A** provides a brief derivation of the Schrodinger-type equation for electrons in gapped graphene. **Appendix B** presents an alternative treatment of the AC electric effect on $\mu_v$.

## II. The valley pair qubit

A single-layer graphene is a gapless, two-dimensional material of hexagonal lattice structure, with a diatomic basis of carbon atoms. The electron energy dispersion simulates that of massless Dirac particles, with $E = \pm v_F|\mathbf{p}|$ ("+" for the conduction band and "−" for the valence band), and the conduction and the valence bands touch at two Dirac points located at K and K', respectively, of the Brilloiun zone.[2,3] "$v_F$" here is the Fermi velocity. The corresponding energy valleys at K and K' are labeled by the isospin index $\tau_v$ ($\tau_v = 1$ for K and $\tau_v = -1$ for K') throughout the work.

The implementation of valley qubits requires the presence of a gap between the conduction and the valence bands. Such a gap may be opened in the graphene layer epitaxially grown on a SiC or BN substrate, due to the substrate-induced asymmetry between the two basis atoms of a unit cell.[28,29] It creates gapped graphene with the corresponding electron dispersion $E = \pm (\Delta^2 + v_F^2 p^2)^{1/2}$ ("+" for the conduction band, and "−" for the valence band) characterized by the energy gap "$2\Delta$" and the effective mass $m^* = \Delta / v_F^2$. In gapped graphene, QDs may be formed via electrostatic modulation of the energy bands in space, where the energy gap is utilized to confine electrons or holes. In the presentation, we shall focus on electrons. (The discussion applies to holes as well.)

Fig. 1 shows the valley qubit system - a DQD structure defined electrostatically, for example, by back gates (not shown here). Gate $V_C$ tunes the potential barrier between the QDs and hence the corresponding interdot tunneling. Gates $V_L$ and $V_R$ are applied to produce DC / AC electric fields across the QDs, for tuning of electron magnetic moments in the QDs, as discussed later.

In the absence of magnetic field, each QD is assumed to accommodate only one quantized energy level. The level is four-fold degenerate, consisting of the quadplet $|\tau_v = \pm 1, \sigma = \pm 1/2\rangle$ (one-electron states), in association with the valley ($\tau_v$) and the spin ($\sigma$) degeneracy. The qubit implementation requires placement of the structure in a tilted, uniform magnetic field, denoted as $\mathbf{B_{total}}$ ($\mathbf{B_{total}} = \mathbf{B_{plane}} + \mathbf{B_{normal}}$). See Fig. 2. Here, $\mathbf{B_{plane}}$ is the in-plane component, and $\mathbf{B_{normal}}$ the component normal to the graphene layer, of the magnetic field. The presence of a tilted magnetic field serves multiple purposes. It removes the redundant spin d.o.f. in the implementation of valley qubits as well as provides the "Larmor precession" of qubits, as shall become clear below.

The magnetic field couples to both the spin and valley d.o.f.s of an electron, according to the following Zeeman-type interaction,

$$H_Z = -g^* \sigma \mu_B |\mathbf{B_{total}}| + \tau_v \mu_v |\mathbf{B_{normal}}|. \tag{II-1}$$

$g^*$ = *electron g-factor*, $\mu_B$ = *Bohr magneton*, and $\mu_v$ is the valley-dependent magnetic moment associated with the orbital motion of an electron in gapped graphene[30]. As expressed in the equation, $\mathbf{B_{total}}$ interacts with the spin d.o.f., giving the usual Zeeman energy "$g^*\sigma\mu_B|\mathbf{B_{total}}|$" and quantizing the spin in the direction of $\mathbf{B_{total}}$, while $\mathbf{B_{normal}}$ couples additionally to the valley d.o.f., giving the corresponding "valley" Zeeman term "$\tau_v\mu_v|\mathbf{B_{normal}}|$". For ½ $g^*$ $\mu_B$ $|\mathbf{B_{total}}|$ > $\mu_v|\mathbf{B_{normal}}|$, the spin splitting leaves $|\tau_v = \pm 1, \sigma = 1/2\rangle$ as the lower valley doublet in each QD. The notations $\{K_L, K_L'\}$ / $\{K_R, K_R'\}$ are used to denote the lower doublets confined in the left / right QDs, respectively. (The spin index is omitted, being fixed at σ = 1/2 here.) These constitute the basis set of one-electron states in forming qubit states, which are two-electron states as explained next.

The valley pair qubit operates in the low energy sector, with the corresponding charge configuration (1,1), meaning that the two electrons are separately confined in the QDs and occupy the one-electron states {$K_L$, $K_L'$} / {$K_R$, $K_R'$}. Virtual electron hopping between the QDs couples the (1,1) configuration to the configuration (0,2) (where the electrons both reside in the same QD), giving rise to the Heisenberg exchange interaction

$$H_J = \frac{1}{4} J \vec{\tau}_L \cdot \vec{\tau}_R, \qquad \text{(II-2)}$$
$$J \sim 4 t_{d-d}^2 / U.$$

Here, J is the exchange coupling, $t_{d-d}$ = interdot tunneling, and U = energy detuning of (0,2) from (1,1). J is tunable via gate $V_C$ (which controls $t_{d-d}$) or back gates (which controls U). $\tau_{L(R)}$ = "*Pauli valley operator*", with

$$\tau_x = \begin{pmatrix} 0 & 1 \\ 1 & 0 \end{pmatrix}, \tau_y = \begin{pmatrix} 0 & -i \\ i & 0 \end{pmatrix}, \tau_z = \begin{pmatrix} 1 & 0 \\ 0 & -1 \end{pmatrix}.$$

The logic 0 / 1 states are represented by the eigenstates of $H_J$, i.e., the "valley singlet $|z_S\rangle$ / triplet $|z_{T0}\rangle$". Here,

$$|z_S\rangle = \frac{1}{\sqrt{2}} (c_{K_L}^+ c_{K_R'}^+ - c_{K_L'}^+ c_{K_R}^+) | vaccum \rangle, \qquad \text{(II-3)}$$
$$|z_{T0}\rangle = \frac{1}{\sqrt{2}} (c_{K_L}^+ c_{K_R'}^+ + c_{K_L'}^+ c_{K_R}^+) | vaccum \rangle,$$

in terms of electron creation operators for the QD-confined states {$K_L$, $K_L'$, $K_R$, $K_R'$}. Linear combinations of {$|z_S\rangle$, $|z_{T0}\rangle$} give

$$|x_+> = c_{K_L}^+ c_{K_R}^+ |vaccum>, \quad |x_-> = c_{K_L'}^+ c_{K_R'}^+ |vaccum>.$$

The qubit state space (denoted as $\Gamma_v$) is expanded by $|z_S>$ and $|z_{T0}>$, and isomorphic to the spin-1/2 state space, e.g., $|z_S> \leftrightarrow |s_z = -1/2>$, $|z_{T0}> \leftrightarrow |s_z = 1/2>$, $|x_-> \leftrightarrow |s_x = -1/2>$, $|x_+> \leftrightarrow |s_x = 1/2>$. The other triplet states,

$$|z_{T+}> = c_{K_L}^+ c_{K_R'}^+ |vaccum>, \quad |z_{T-}> = c_{K_L'}^+ c_{K_R}^+ |vaccum>,$$

are outside $\Gamma_v$ and not needed for quantum computing.

Eqns. (II-1) and (II-2) determine the energy levels of various two-electron states. See Fig. 3. $\mathbf{B_{normal}}$ splits $|z_{T\pm}>$ away from $\{|z_S>, |z_{T0}>\}$, by $(\mu_{vL} + \mu_{vR})|\mathbf{B_{normal}}|$. $|z_S>$ and $|z_{T0}>$ are separated by J, while $|x_->$ and $|x_+>$ by $2(\mu_{vL} - \mu_{vR})|\mathbf{B_{normal}}|$. Here, $\mu_{vL} / \mu_{vR}$ are the valley magnetic moments in the left / right QDs, respectively. Overall, the qubit system is described by the following effective Hamiltonian in the reduced space $\Gamma_v$,

$$H_{eff} = (\mu_{vL} - \mu_{vR}) B_{normal} \tau_x + \frac{J}{2} \tau_z \qquad (II-4)$$

in the basis of $\{|z_S>, |z_{T0}>\}$.

$H_{eff}$ governs the time evolution of the qubit state. The $\tau_x$ part generates a rotation $\check{R}_x(\theta_x = \Omega_x t_x)$ about the x-axis (of the Bloch sphere) when it is applied for the time $t_x$, where $\Omega_x$ is the "Larmor frequency", e.g., $\Omega_x = 2(\mu_{vL}-\mu_{vR}) B_{normal}/\hbar$. The $\tau_z$ part generates a rotation $\check{R}_z(\theta_z = \Omega_z t_z)$ about the z-axis ($\Omega_z = J/\hbar$, and $t_z$ = corresponding time). $\check{R}_x$ and $\check{R}_z$, together, allow the manipulation of a single qubit. See Fig. 4.

According to Eqn. (II-4), a gradient in the magnetic moment, i.e., $\mu_{vL} \neq \mu_{vR}$, is a required condition for the manipulation. Because the magnetic moment is energy-dependent, it can be modulated by tuning the electron energy level in the QD (as discussed in **Sec. V**). Therefore, a structural differentiation in the QDs may provide the required gradient. There are also controllable, electric means of modulating $\mu_{vL}$ / $\mu_{vR}$ in DC or AC modes via back gates or gates $V_L$ / $V_R$, in order to furnish the gradient (the latter being discussed in **Sec. VI**). In the DC mode, ($\mu_{vL}$ - $\mu_{vR}$) is time-independent, and an initial qubit state, e.g., $|\Phi(t=0)\rangle = |z_S\rangle$, undergoes the following rotation

$$|\Phi(t)\rangle = \hat{R}_n(\theta_n^{(DC)})|z_S\rangle, \tag{II-5}$$

$$\hat{R}_n(\theta_n^{(DC)}) = \exp\left(-\frac{i}{2}\vec{\tau}\cdot\vec{\theta}_n^{(DC)}\right),$$

$$\vec{\theta}_n^{(DC)} = \frac{t}{\hbar}\left(2(\mu_{vL}-\mu_{vR})B_{normal}, 0, J\right),$$

where the direction and the magnitude of $\theta_n^{(DC)}$ determine the axis and the angle of rotation, respectively. Tuning of the ratio, ($\mu_{vL}$-$\mu_{vR}$) : J, by $V_L$, $V_R$, and $V_C$ (or back gates) determines the direction of $\theta_n^{(DC)}$, and tuning of the time length, t, determines the magnitude of $\theta_n^{(DC)}$. On the other hand, the manipulation can also be performed in the AC mode, where ($\mu_{vL}$ - $\mu_{vR}$) varies periodically in time, as discussed in **Sec. VI**. It allows the application of qubit resonance techniques to qubit manipulation. Although the application has initially been suggested to be performed via periodic modulation of the exchange coupling J,[10] the alternative provided here– periodic modulation of ($\mu_{vL}$ - $\mu_{vR}$) adds another dimension to it.

This qubit implementation is the analogue of the spin pair scheme[9-13]. It shares the distinctive advantages provided in the scheme, e.g., scalability and fault-tolerance, and the method developed in the scheme for initialization / readout / qugate operation[12] may be adapted here.

As Eqn. (II-4) forms the basis of qubit manipulation, it constitutes the core of discussion in **Secs. III~VI**. These sections supplement the argument leading to Eqn. (II-4), estimate the parameters (e.g., $\mu_{vL}$, $\mu_{vR}$, and J) in the equation, discuss ways to control the parameters, and the qubit manipulation based on the equation.

## III. The quantum mechanical description

It is obvious from **Sec. II** that tuning of the magnetic moment ($\mu_{vL}$ and $\mu_{vR}$) is an important part of the qubit manipulation. In order to prepare the background for the discussion of tuning, this section describes the quantum mechanics of a graphene QD in the presence of both an in-plane electric field and a normal magnetic field, $B_{normal}$. An electron in the case satisfies the following Dirac-type equation,[3]

$$H_D \phi_D = E \phi_D, \qquad \text{(III-1)}$$

$$H_D = H_0(\tau_v) + H_{electric} + H_{magnetic}(\tau_v), \quad \phi_D = \begin{pmatrix} \phi_A \\ \phi_B \end{pmatrix},$$

$$H_0(\tau_v) = \begin{pmatrix} \Delta & v_F(\hat{p}_x - i\tau_v \hat{p}_y) \\ v_F(\hat{p}_x + i\tau_v \hat{p}_y) & -\Delta \end{pmatrix},$$

$$H_{electric} = \begin{pmatrix} V & 0 \\ 0 & V \end{pmatrix}, \quad H_{magnetic}(\tau_v) = ev_F \begin{pmatrix} 0 & A_x - i\tau_v A_y \\ A_x + i\tau_v A_y & 0 \end{pmatrix},$$

$$V = V_{QD} + V_\varepsilon.$$

Here, "$H_D$" is the Dirac Hamiltonian, "$\Phi_D$" is the two-component Dirac wave function, $V_{QD}$ = confining QD potential energy, $V_\varepsilon$ = external electric potential energy, and ($A_x$, $A_y$) is the vector potential in association with the magnetic field.

For the qubit implementation, we consider the case where E ~ $\Delta$ (the "non-relativistic limit"). In this limit, Eqn. (III-1) is reduced to the following Schrodinger-type equation [correct to O($B_{normal}$) and the 1$^{st}$-order R.C.],

$$H\phi = E\phi, \qquad \text{(III-2)}$$
$$H = H^{(0)} + H^{(1)}.$$

$H^{(0)}$ is the "nonrelativistic part", with

$$H^{(0)} = \frac{\vec{\pi}^2}{2m^*} + V + \tau_v \mu_{v0} B_{normal},$$

and $H^{(1)}$ is the "1$^{st}$-order R.C.", with

$$H^{(1)} = \tau_v \frac{\hbar}{4m^*\Delta} [(\nabla V) \times \vec{\pi}]_{normal} \quad (VOI)$$

$$- \frac{1}{2\Delta} \left( \frac{\vec{\pi}^2}{2m^*} + \tau_v \mu_{v0} B_{normal} \right)^2 - \frac{1}{8m^*\Delta} (\vec{p}^2 V),$$

$$\vec{\pi} = \vec{p} + e\vec{A}, \quad \mu_{v0} = \frac{e\hbar}{2m^*}.$$

"H" is the Schrodinger Hamiltonian, and "Φ" denotes the one-component Schrodinger wave function. The electron energy E is now defined with respect to the conduction band edge. $H^{(0)}$ describes the motion of a charged particle in the magnetic field **B$_{normal}$**. The last term "$\tau_v$ $\mu_{v0}$ B$_{normal}$" in $H^{(0)}$ gives rise to the valley splitting in the field, with $\mu_{v0}$ = eℏ/2m$^*$ being the nonrelativistic valley magnetic moment, the analogue of Bohr spin magneton. The 1$^{st}$ term in $H^{(1)}$ is called the valley-orbit interaction (VOI),[31] the analogue of spin-orbit interaction. The 2$^{nd}$ term is the usual 1$^{st}$-order relativistic correction to the non-relativistic kinetic energy. The 3$^{rd}$ term is the Darwin's term. Eqn. (III-2) is derived in **Appendix A**. Eqns. (III-1) and (III-2) shall be utilized in **Sec. V**.

## IV. The exchange coupling J

The tuning of the exchange coupling constitutes an important part of qubit manipulation in the quantum dot approach.[8,10-13] This is also true in the case of valley pair qubits. In this section, the coupling is evaluated for near-gap electrons (i.e., the non-relativistic limit). We consider the double graphene QDs (with size ~ L), as shown in Fig. 1, and estimate J using the formula

$$J \sim 4t_{d-d}^2/U \tag{II-2}$$

We assume that the two QDs are identical, and there is only one confined level in each QD, with energy E (relative to the conduction band edge). U here reduces to the on-site Coulomb energy. The depth of QD potential is denoted as $V_0$ and taken to be of the order of E. Specifically, we use $V_0 = 2E$. In the non-relativistic limit, where the Schrodinger type equation (III-2) applies, we estimate roughly

$$t_{d-d} \sim O(1) E \exp(-2\alpha W), \tag{IV-2}$$
$$\alpha = \frac{\sqrt{2m^*(V_0 - E)}}{\hbar}.$$

Here, W = *interdot potential barrier width*. Using $\Delta \sim 0.14\text{eV}^{28}$, $v_F = 10^6$ m/s, L ~ 350Å, E ~ 30 meV, U ~ 10meV, and W ~ 0.5 L in Eqns. (IV-1) and (IV-2), we obtain J ~ O(1) meV. Lower values of J can be obtained by increasing W, the interdot barrier height (via gate $V_C$), or the detuning U (via back gates)[12,13].

## V. The valley splitting and the valley magnetic moment in a QD

This section discusses the valley splitting in a QD, in the presence of a normal magnetic field **B**$_{\text{normal}}$, as well as the valley magnetic moment $\mu_v$. The result shall be utilized in the discussion of magnetic moment tuning in **Sec. VI**. Throughout the work, we assume the weak magnetic field limit, e.g.,

$$l_{orbital} \ll l_B,$$
$$l_{orbital} \sim L, \; l_B = \sqrt{\frac{\hbar}{eB_{normal}}}.$$

$l_{orbital}$ is the electron orbital size, L the QD size, and $l_B$ the magnetic length. We shall neglect the terms which are $O(B_{normal}^2)$ in the calculation. Moreover, in the regime considered here, QD size confinement dominates over Landau level quantization, allowing the discussion of magnetic effects here to be carried out within the perturbation theory.

General discussion of valley splitting is given first, based on the Dirac Eqn. (III-1).[32] We begin with the case of gapless graphene, i.e., $\Delta = 0$. It is straightforward to verify that if $(\Phi_A, \Phi_B)$ is a solution to Eqn. (III-1) for $\tau_v = 1$, then in association with it, there is a degenerate solution, $(\Phi_B, \Phi_A)$, with $\tau_v = -1$. Therefore, in order to produce a valley splitting, the condition $\Delta \neq 0$ is required. We turn to the case of gapped graphene below.

The valley splitting in the case is analyzed in the framework of perturbation theory. Again, Eqn. (III-1) is utilized for a general discussion first. For $B_{normal} = 0$, $H_{magnetic}$ vanishes and Eqn. (III-1) reduces to

$$[H_0(\tau_v) + H_{electric}]\phi_D^{(0)}(\tau_v) = E^{(0)}\phi_D^{(0)}(\tau_v). \tag{V-1}$$

Owing to the time-reversal symmetry, the states for $\tau_v = \pm 1$ are degenerate. Specifically, one can verify that if $(\Phi_A^{(0)}, \Phi_B^{(0)})$ is a solution to Eqn. (V-1) for $\tau_v = 1$, then $(-\Phi_B^{(0)*}, \Phi_A^{(0)*})$ is a solution with the same energy, for $\tau_v = -1$, i.e.,

$$\left(\phi_A^{(0)}(x,y;\tau_v), \phi_B^{(0)}(x,y;\tau_v)\right) = \left(-\phi_B^{(0)*}(x,y;-\tau_v), \phi_A^{(0)*}(x,y;-\tau_v)\right) \tag{V-2}$$

For $B_{normal} \neq 0$, we treat $H_{magnetic}$ as a perturbation. [We choose the asymmetric gauge $(A_x, A_y) = (0, B_{normal} x)$ to simplify the discussion here.] It gives the following 1$^{st}$-order energy correction (linear in $B_{normal}$)

$$E_Z(\tau_v) = <\phi_D^{(0)}(\tau_v)|H_{magnetic}(\tau_v)|\phi_D^{(0)}(\tau_v)> = 2ev_F\tau_v\text{Imag}\left[<\phi_A^{(0)}(\tau_v)|A_y|\phi_B^{(0)}(\tau_v)>\right] \tag{V-3}$$

Using Eqn. (V-2), one can show that $\text{Imag}\left[<\phi_A^{(0)}(\tau_v)|A_y|\phi_B^{(0)}(\tau_v)>\right]$ here is actually $\tau_v$-independent. Thereby, $E_Z(\tau_v) \propto \tau_v$, carrying opposite signs for the $\tau_v = \pm 1$ states and showing "Zeeman valley splitting". Moreover, the splitting size is generally QD structure-dependent, as it depends on the wave function of the quantized state.

Now, we proceed to a more detailed discussion of valley splitting, based on the Schrodinger-type equation (III-2). We consider specifically a rectangular QD, and write (correct to $O(B_{normal})$)

$$H \simeq H_{B\tau}^{(0)} + H_{B0}^{(0)} + H_{0\tau}^{(1)} + H_{B\tau}^{(1)} + H_{rest}, \qquad (V-4)$$

$$H_{B\tau}^{(0)} = \frac{\vec{p}^2}{2m^*} + V + \tau_v \mu_{v0} B_{normal}, \quad H_{B0}^{(0)} = \frac{e\vec{A} \cdot \vec{p}}{m^*},$$

$$H_{0\tau}^{(1)} = \tau_v \frac{\hbar}{4m^*\Delta}\left[(\nabla V) \times \vec{p}\right]_{normal}, \quad H_{B\tau}^{(1)} = -\tau_v \mu_{v0}\left(B_{normal} \frac{\vec{p}^2}{2m^*\Delta} + \frac{1}{2\Delta}\left[\vec{A}\times(\nabla V)\right]_{normal}\right).$$

Here, $V = V_{QD} = \frac{1}{2} m^*(w_x^2 x^2 + w_y^2 y^2)$ (for a rectangular QD), with $O(w_x) \sim O(w_y)$. H has been expressed in a form convenient for the perturbative calculation. Each term in H is labeled with two subscripts, with the 1st one ("B" or "0") denoting whether it is $B_{normal}$-dependent, and the 2nd one ("τ" or "0") denoting whether it is $\tau_v$-dependent. $H_{B\tau}^{(0)}$ describes a 2D anisotropic simple harmonic oscillator with valley splitting. The corresponding oscillator function is denoted as $\Phi_{nm}^{(0)}$ ((n, m) being the 2D harmonic oscillator indices). The other terms in Eqn. (V-4) are treated as the perturbation. These terms (except $H_{rest}$) lead to *the energy correction linear in both $B_{normal}$ and $\tau_v$* (i.e., the Zeeman energy $E_Z$). $H_{rest}$ is both $\tau_v$- and $B_{normal}$- independent, and irrelevant for the calculation of valley splitting.

The perturbation theory gives the following Zeeman energy and valley magnetic moment,

$$E_Z(\tau_v) \approx \tau_v \mu_{v0} B_{normal} + <\phi_{00}^{(0)} | H_{B\tau}^{(1)} | \phi_{00}^{(0)}> \tag{V-5}$$

$$+2\text{Real}\left[<\phi_{00}^{(0)} | H_{B0}^{(0)} | \phi_{11}^{(0)}> \frac{1}{\hbar(-w_x - w_y)} <\phi_{11}^{(0)} | H_{0\tau}^{(1)} | \phi_{00}^{(0)}>\right]$$

$$= \tau_v \mu_v B_{normal},$$

$$\mu_v = \mu_{v0}\left[1 - \frac{\hbar}{2\Delta} \frac{w_x w_y}{(w_x + w_y)}\right].$$

The 2nd term in $\mu_v$ derives from the 1st-order R.C..

## VI.    $\mu_v$-tuning in DC/AC modes and qubit manipulation

As Eqn. (V-5) shows, $\mu_v$ depends on $w_x$ and $w_y$ – parameters in the QD potential $V_{QD}$. It suggests a way to tune $\mu_v$ via modulation of the potential. We investigate electric means of the modulation below, in either DC or AC modes.

Firstly, we consider a QD in the presence of a DC uniform electric field ($\varepsilon_x^{(DC)}$ in the x-direction) as well as the normal, uniform magnetic field, $\mathbf{B}_{normal}$, where $V_{QD}$ consists of a dominant quadratic potential and a weak anharmonic one. Two cases are analyzed. In one case, the anharmonic term is cubic ($x^3$). In the other, it is quadric ($x^4$). The presence of the anharmonic term is required because, without it, the electric field would just shift the electron equilibrium position without altering the physics.

We consider the cubic case now. Let

$$V = V_{QD} + e\varepsilon_x^{(DC)} x,$$
$$V_{QD} = V_2 + V_3,$$
$$V_2 = \frac{1}{2} m^* w_0^2 (x^2 + y^2), \quad V_3 = \frac{1}{3} k_{3x} m^* w_0^2 x^3.$$

Here, $k_{3x}$ characterizes the strength of $V_3$. We assume that both $k_{3x}$ and $\varepsilon_x^{(DC)}$ are weak ($k_{3x} \ll (m^* w_0/\hbar)^{1/2}$, $e\varepsilon_x^{(DC)} \ll (\hbar m^* w_0^3)^{1/2}$), and derive the effect of $\varepsilon_x^{(DC)}$ on the valley splitting. We introduce $x_\varepsilon^{(DC)} \equiv -e\varepsilon_x^{(DC)}/m^* w_0^2$, the shift in electron equilibrium position due to the electric field, and write

$$V \approx \frac{1}{2} m^* w_0^2 \left[ (x - x_\varepsilon^{(DC)})^2 + y^2 \right] + \frac{1}{3} k_{3x} m^* w_0^2 (x - x_\varepsilon^{(DC)})^3$$
$$+ k_{3x} m^* w_0^2 x_\varepsilon^{(DC)} (x - x_\varepsilon^{(DC)})^2,$$
$$\xrightarrow{x - x_\varepsilon^{(DC)} \to x} \frac{1}{2} m^* w_0^2 (x^2 + y^2) + \frac{1}{3} k_{3x} m^* w_0^2 x^3 + k_{3x} m^* w_0^2 x_\varepsilon^{(DC)} x^2,$$

correct up to $O(x_\varepsilon^{(DC)})$. Here, a change of coordinates $x - x_\varepsilon^{(DC)} \to x$ has been made. The above result shows that the DC field modifies the quadratic potential as follows,

$$\frac{1}{2} m^* w_0^2 (x^2 + y^2) \to \frac{1}{2} m^* (w_x^2 x^2 + w_0^2 y^2),$$
$$w_x^2 = w_0^2 (1 + 2 k_{3x} x_\varepsilon^{(DC)}),$$

giving it a slight anisotropy. Using the result in Eqn. (V-5) for a rectangular QD, we obtain the $\varepsilon_x^{(DC)}$-induced Zeeman energy,

$$\delta E_Z^{(DC)}(\tau_v) \approx -\frac{1}{8} \tau_v \mu_{v0} B_{normal} \left( k_{3x} x_\varepsilon^{(DC)} \right) \frac{\hbar w_0}{\Delta}. \tag{VI-1}$$

Correspondingly, it gives the $\varepsilon_x^{(DC)}$-induced valley magnetic moment tuning,

$$\delta \mu_v^{(DC)} \approx -\mu_{v0} \frac{1}{8} \left( k_{3x} x_\varepsilon^{(DC)} \right) \frac{\hbar w_0}{\Delta} \tag{VI-2}$$

The results in Eqns. (VI-1) and (VI-2) both are linear in $\hbar w_0/\Delta$, the ratio of the ground state energy to the "rest mass" energy, showing that these are the 1st -order "relativistic effects".

Next, we consider the quadric case,

$$V = V_{QD} + e\varepsilon_x^{(DC)} x,$$
$$V_{QD} = V_2 + V_4,$$
$$V_2 = \frac{1}{2} m^* w_0^2 (x^2 + y^2), \quad V_4 = \frac{1}{4} k_{4x} m^* w_0^2 x^4.$$

Here, $k_{4x}$ characterizes the strength of $V_4$. Note that the energy eigenvalue in this case is an even function, i.e., $E(\tau_v; \varepsilon_x^{(DC)}) = E(\tau_v; -\varepsilon_x^{(DC)})$. [It can be shown that, if $\Phi(x,y;\tau_v)$ is an eigenstate (in the field "$\varepsilon_x^{(DC)}$") with the energy $E(\tau_v; \varepsilon_x^{(DC)})$, then $\Phi^*(-x,y;\tau_v)$ is a solution (in the field "$-\varepsilon_x^{(DC)}$") with the same energy.] This result is used below. We also note that, $V_2+V_4$ with $k_{4x} < 0$ describes a parabolic confining potential with a cut-off in the x-direction. (One can also add a similar $y^4$ term to cut the potential off in the y-direction, without affecting the discussion in this section.) *Therefore, the presence of $V_4$ is a realistic assumption.*

We assume that both $k_{4x}$ and $\varepsilon_x^{(DC)}$ are weak ($k_{4x} \ll m^* w_0 / \hbar$, $e\varepsilon_x^{(DC)} \ll (\hbar m^* w_0^3)^{1/2}$), and derive the effect of $\varepsilon_x^{(DC)}$ on the valley splitting. We write

$$V \approx \frac{1}{2} m^* w_0^2 \left[ (x - x_\varepsilon^{(DC)})^2 + y^2 \right] + \frac{1}{4} k_{4x} m^* w_0^2 (x - x_\varepsilon^{(DC)})^4$$
$$+ k_{4x} m^* w_0^2 x_\varepsilon^{(DC)} (x - x_\varepsilon^{(DC)})^3 + \frac{3}{2} k_{4x} m^* w_0^2 x_\varepsilon^{(DC)2} (x - x_\varepsilon^{(DC)})^2,$$

$$\xrightarrow{x - x_\varepsilon^{(DC)} \to x} \frac{1}{2} m^* w_0^2 (x^2 + y^2) + \frac{1}{4} k_{4x} m^* w_0^2 x^4 + k_{4x} m^* w_0^2 x_\varepsilon^{(DC)} x^3 + \frac{3}{2} k_{4x} m^* w_0^2 x_\varepsilon^{(DC)2} x^2,$$

$$x_\varepsilon^{(DC)} = -e\varepsilon_x^{(DC)} / m^* w_0^2,$$

correct up to $O(x_\varepsilon^{(DC)2})$. We further drop the linear-in-$x_\varepsilon^{(DC)}$ cubic potential term, "$k_{4x} m^* w_0^2 x_\varepsilon^{(DC)} x^3$", from V. Since $E(\tau_v; \varepsilon_x^{(DC)}) = E(\tau_v; -\varepsilon_x^{(DC)})$, the contribution of this term to the energy cannot be linear. Instead, it is $O(k_{4x}^2 x_\varepsilon^{(DC)2})$ or of a higher order, which is negligible in comparison to the result that will be derived below. Thereby, we write

$$V \approx \frac{1}{2}m^*(w_x^2 x^2 + w_0^2 y^2) + \frac{1}{4}k_{4x}m^*w_0^2 x^4,$$
$$w_x^2 = w_0^2(1 + 3k_{4x}x_\varepsilon^{(DC)2}).$$

The DC field again brings a slight anisotropy to the harmonic part of potential. Using the result in Eqn. (V-5) for a rectangular QD, we obtain the $\varepsilon_x^{(DC)}$-induced Zeeman energy,

$$\delta E_Z^{(DC)}(\tau_v) \approx -\frac{3}{16}\tau_v \mu_{v0} B_{normal} \left(k_{4x}x_\varepsilon^{(DC)2}\right) \frac{\hbar w_0}{\Delta}. \tag{VI-3}$$

This corresponds to the $\varepsilon_x^{(DC)}$-induced valley magnetic moment tuning

$$\delta\mu_v^{(DC)} \approx -\mu_{v0}\frac{3}{16}\left(k_{4x}x_\varepsilon^{(DC)2}\right)\frac{\hbar w_0}{\Delta} \tag{VI-4}$$

In either the cubic or the quadric case, the $\mu_v$-tuning can be utilized to create the required asymmetry ($\mu_{vL} \neq \mu_{vR}$) in the DQD structure of the qubit, and generate a qubit state rotation, $\check{R}_x$, about the x-axis (of the Bloch sphere). Specifically, we consider the case of a symmetric DQD structure, where $\mu_{vL} = \mu_{vR}$. If a DC electric field is applied on the left QD, for the time $t_x$, a qubit state, $|\Phi\rangle$, will undergo the following rotation (according to Eqn. (II-5))

$$|\Phi\rangle \rightarrow \hat{R}_x(\theta_x^{(DC)})|\Phi\rangle, \tag{VI-5}$$
$$\hat{R}_x(\theta_x^{(DC)}) = \exp\left(-i\tau_x \frac{\theta_x^{(DC)}}{2}\right),$$
$$\theta_x^{(DC)} = \Omega_x t_x, \quad \Omega_x = \frac{2}{\hbar}\delta\mu_v^{(DC)} B_{normal}.$$

Here, we have set the exchange coupling $J \sim 0$ in Eqn. (II-5), and replaced $\mu_{vL}-\mu_{vR}$ in the equation with $\delta\mu_v^{(DC)}$ (induced by the field in the left QD). With the parameters $B_{normal}=100mT$, $k_{4x}=L^{-2}$ (L = QD size), $x_\varepsilon^{(DC)}/L= 0.2$, $\hbar w_0/\Delta=0.2$, $\Delta=0.14eV$,[28] Eqns. (VI-4) and (VI-5) give $\Omega_x$ (the Larmor frequency for $\check{R}_x$) ~ O(ns$^{-1}$), in the typical range currently envisioned in the QD approach. This estimate of manipulation speed remains valid when a finite J is included, since, as discussed in **Sec. IV**, J can reach O(meV), with the corresponding $\Omega_z \gg$ O(ns$^{-1}$) ($\Omega_z$ = J/ ℏ, the Larmor frequency for $\check{R}_z$).

AC mode of $\mu_v$-tuning can also be achieved with an AC field superimposed on the DC field, e.g.,

$$\varepsilon_x = \varepsilon_x^{(DC)} + \varepsilon_x^{(AC)} \sin(w_{AC}t).$$

Under the adiabatic condition, e.g., $w_{AC} \ll w_0$, where the AC field does not cause any transition between QD energy levels, the following additional tuning ($\delta\mu_v^{(AC)}$) is obtained, namely, for weak $\varepsilon_x^{(AC)}$ ($e\varepsilon_x^{(AC)} \ll (\hbar m^* w_0^3)^{1/2}$ in the cubic case, and $\varepsilon_x^{(AC)} \ll \varepsilon_x^{(DC)}$ in the quadric case),

$$\delta\mu_v = \delta\mu_v^{(DC)} + \delta\mu_v^{(AC)}, \tag{VI-6}$$

$$\delta\mu_v^{(AC)} \approx -\frac{1}{8}\mu_{v0}\left(k_{3x}x_\varepsilon^{(AC)}\right)\sin(w_{AC}t)\frac{\hbar w_0}{\Delta} \quad \text{(cubic case)} \tag{VI-7}$$

$$\delta\mu_v^{(AC)} \approx -\frac{3}{8}\mu_{v0}\left(k_{4x}x_\varepsilon^{(DC)}x_\varepsilon^{(AC)}\right)\sin(w_{AC}t)\frac{\hbar w_0}{\Delta}, \quad \text{(quadric case)} \tag{VI-8}$$

$$x_\varepsilon^{(AC)} \equiv -\frac{e\varepsilon_x^{(AC)}}{m^* w_0^2},$$

correct to the order of $\varepsilon_x^{(AC)}$. This result can be derived by a straightforward application of Eqns. (VI-2) and (VI-4), where $x_\varepsilon^{(DC)}$ is replaced by $x_\varepsilon^{(DC)} + x_\varepsilon^{(AC)}\sin(w_{AC}t)$.

We mentioned that the AC mode allows the qubit manipulation via qubit resonance techniques. A simplified picture of the manipulation is given here which involves alternating $\check{R}_x$ and $\check{R}_z$, as follows. We assume that both the QDs are subject to DC fields, and gates $V_L$ and $V_R$ have been tuned such that the DC part of ($\mu_{vL}$-$\mu_{vR}$), ($\mu_{vL}$-$\mu_{vR}$)$^{(DC)}$, vanishes. Moreover, we assume that only the left QD is subject to an AC field, and write the AC part of ($\mu_{vL}$-$\mu_{vR}$), ($\mu_{vL}$-$\mu_{vR}$)$^{(AC)} = \delta\mu_v^{(AC)}$. For one half of the AC cycle, the AC field induces the transformation $|\Phi\rangle \rightarrow \check{R}_x(\theta_x^{(AC)})|\Phi\rangle$, with

$$\theta_x^{(AC)} = \frac{2}{\hbar} \int_0^{\pi/w_{AC}} \delta\mu_v^{(AC)} B_{normal} dt$$

$$\theta_x^{(AC)} = -\frac{1}{2} k_{3x} x_\varepsilon^{(AC)} \frac{\mu_{v0} B_{normal}}{\hbar w_{AC}} \frac{\hbar w_0}{\Delta} \qquad \text{(cubic case)} \qquad (VI\text{-}9)$$

$$\theta_x^{(AC)} = -\frac{3}{2} k_{4x} x_\varepsilon^{(DC)} x_\varepsilon^{(AC)} \frac{\mu_{v0} B_{normal}}{\hbar w_{AC}} \frac{\hbar w_0}{\Delta} \qquad \text{(quadric case)} \qquad (VI\text{-}10)$$

In the other half of the cycle, it rotates the qubit state through "$-\theta_x^{(AC)}$". In general, the qubit may be manipulated, e.g., in the alternating sequence $\check{R}_x(\theta_x^{(AC)}) \rightarrow \check{R}_z(\theta_z=\pi) \rightarrow \check{R}_x(-\theta_x^{(AC)}) \rightarrow \check{R}_z(\theta_z=\pi) \rightarrow \ldots \check{R}_z(\theta_z^{(target)}+\pi/2)$, into a target state ($\theta_z^{(target)} = $ *target state longitude*). See Fig. 5. Here, we have assumed 1) J = 0 when $\check{R}_x$ is acting on the qubit state, and 3) between the half cycles, the AC field is turned off and J is turned on, for the time length $\pi\hbar/J$, when $\check{R}_z$ is acting on the qubit state. Under the foregoing assumptions 1) and 2), rotations about the x- and z- axes are basically decoupled.

In the quadric case, the discussion of AC mode for $\varepsilon_x^{(AC)} \ll \varepsilon_x^{(DC)}$ can be extended to the regime where $\varepsilon_x^{(AC)} \sim \varepsilon_x^{(DC)}$. In order to see this, we write

$$\delta\mu_v = \delta\mu_v^{(DC)} + \delta\mu_v^{(AC)} + \delta\mu_v^{(AC)'},$$

$$\delta\mu_v^{(AC)'} = -\frac{3}{16} \mu_{v0} \left(k_{4x} x_\varepsilon^{(AC)2}\right) \sin^2(w_{AC} t) \frac{\hbar w_0}{\Delta},$$

correct to the order of $(\varepsilon_x^{(AC)})^2$. Therefore, an additional term, $\delta\mu_v^{(AC)'}$, arises in the regime. It leads to a rotation through the extra angle

$$\theta_x^{(AC)'} = \frac{2}{\hbar} \int_0^{\pi/w_{AC}} \delta\mu_v^{(AC)'} B_{normal} dt,$$

in each half of the AC cycle. Being quadratic in $\varepsilon_x^{(AC)}$, the sign of $\theta_x^{(AC)'}$ does not flip during the whole AC cycle. However, this additional rotation is nullified in the alternating sequence of operations, since

$$\check{R}_x(\theta_x^{(AC)'}) \cdot \check{R}_z(\theta_z=\pi) \cdot \check{R}_x(\theta_x^{(AC)'}) = \check{R}_z(\theta_z=\pi),$$

as can be verified.

**Appendix B** provides an alternative treatment of the AC electric effect on $\mu_v$ employing a moving reference frame.

## VII. The phonon-mediated valley relaxation in a QD, and singlet-triplet transition in a QD or DQD

A qubit state may relax or dephase. It constrains the time allowed for qubit manipulation. This issue is discussed below in the case of valley pair qubits.

Firstly, we estimate the valley relaxation time in a QD, in the presence of the normal magnetic field, **B**$_{normal}$. We take L (the QD size) >> Å. The QD-confined states are denoted as |K$_{QD}$> and |K$_{QD}$'>, corresponding to the two energy valleys K and K', respectively. The valley splitting is written as $\delta E_{KK'}$ (~ $2\mu_{v0}B_{normal}$).

We write |K$_{QD}$> and |K$_{QD}$'> as follows,

$$|K_{QD}> = c_{QD}^{(K)+}|vaccum>, \quad |K'_{QD}> = c_{QD}^{(K')+}|vaccum>, \qquad (\text{VII-1})$$

$$K_{QD}(\vec{r}) = <\vec{r}|K_{QD}> \approx \phi(\vec{r}-\vec{R}_{QD};\tau_v=1)e^{i\vec{K}\cdot\vec{r}}u_K,$$

$$K'_{QD}(\vec{r}) = <\vec{r}|K'_{QD}> \approx \phi(\vec{r}-\vec{R}_{QD};\tau_v=-1)e^{i\vec{K}'\cdot\vec{r}}u_{K'},$$

$$\vec{K}(\vec{K}') = \text{wave vector at K(K') point},$$

$$\vec{R}_{QD} = QD\ center\ postion,$$

$$u_{K(K')} = Bloch\ cell\text{-}periodic\ function.$$

$c_{QD}^{(K)+}$ and $c_{QD}^{(K')+}$ are electron creation operators. $\Phi$ = *envelope function*, taken to be smoothly varying, with a width L (due to the QD confinement).

The valley relaxation process occurs via the quantum state flip $|K_{QD}> \leftrightarrow |K_{QD}'>$. The process involves the intervalley scattering between K and K' and, hence, a corresponding large wave vector transfer ($\delta k = |\vec{K}-\vec{K}'| \sim $ Å$^{-1}$). It may occur easily in the presence of a short-range impurity potential which scatters the electron and provides the required wave vector difference.[33] In this work, however, we consider the clean limit where the structure is free of such impurities, and $\delta k$ must be furnished by the confining QD potential. In addition, because of the energy mismatch, $\delta E_{KK'}$, between $|K_{QD}>$ and $|K_{QD}'>$, the electron-phonon (EP) interaction also participates in the transition, making up for the energy difference. (This is similar to the phonon-mediated spin relaxation in semiconductor QDs.[20,21]) Our calculation below is an estimate of the phonon-mediated valley-flip rate in a QD.

The EP interaction is modeled by the deformation potential coupling involving acoustic phonons,

$$H_{EP} = \sum_Q \sum_{v=K,K'} M_{EP}^{(v)}(Q)(b_{-Q}+b_Q^+)c_{QD}^{(v)+}c_{QD}^{(v)}, \qquad (\text{VII-2})$$

with the matrix elements

$$M_{EP}^{(K)}(Q) = D|Q|\sqrt{\frac{\hbar}{2\rho_a A w_Q}} < K_{QD} | e^{i\bar{Q}\cdot\bar{r}} | K_{QD} >, \tag{VII-3}$$

$$M_{EP}^{(K')}(Q) = D|Q|\sqrt{\frac{\hbar}{2\rho_a A w_Q}} < K_{QD}' | e^{i\bar{Q}\cdot\bar{r}} | K_{QD}' >.$$

b and $b^+$ are phonon creation and annihilation operators, D = *deformation potential constant*, Q = *phonon wave vector*, $\rho_a$ = *mass density*, A = *system area*, $w_Q = c_s Q$ (acoustic phonon dispersion), and $c_s$ = *sound velocity*.

The QD potential is written below

$$V_{QD} \approx M_{QD}^{(KK')} c_{QD}^{(K)+} c_{QD}^{(K')} + h.c., \tag{VII-4}$$

$$M_{QD}^{(KK')} = < K_{QD} | V_{QD} | K_{QD}' >, \quad M_{QD}^{(K'K)} = < K_{QD}' | V_{QD} | K_{QD} >.$$

We take the depth of $V_{QD}$ to be $V_0$. We estimate

$$|M_{QD}^{(KK')}| \approx |M_{QD}^{(K'K)}| \approx M_{QD}, \tag{VII-5}$$

$$M_{QD} = \frac{V_0}{1+(L\delta k)^2}.$$

The Lorentzian function with a width 1/L is used here, and below, for the Fourier transform of a $\bar{r}$-space function with spatial width ~ L.

The phonon-mediated valley flip, $|K_{QD}> \leftrightarrow |K_{QD}'>$, is a 2[nd]-order process involving the following products of matrix elements,

$$< K_{QD}', Q | H_{EP} | K_{QD}' >< K_{QD}' | V_{QD} | K_{QD} >, \tag{VII-6}$$
$$< K_{QD}', Q | V_{QD} | K_{QD}, Q >< K_{QD}, Q | H_{EP} | K_{QD} >,$$
(both with phonon emission),

or

$$< K_{QD} | H_{EP} | K_{QD}, Q >< K_{QD}, Q | V_{QD} | K_{QD}', Q >, \quad \text{(VII-7)}$$
$$< K_{QD} | V_{QD} | K_{QD}' >< K_{QD}' | H_{EP} | K_{QD}', Q >,$$
(both with phonon absorption).

Composite electron-phonon states appear here. For instance, $|K_{QD}, Q>$ denotes the one-electron ($K_{QD}$)- one phonon (Q) state.

For the transition $|K_{QD}> \leftrightarrow |K_{QD}'>$, Fermi's golden rule gives the rate

$$\frac{1}{\tau_{K' \to K}} = \frac{2\pi}{\hbar} \sum_Q n_Q |M_{K' \to K}|^2 \delta(|\delta E_{KK'}| - \hbar w_Q) \quad \text{(with phonon absorption)}, \quad \text{(VII-8)}$$

$$\frac{1}{\tau_{K \to K'}} = \frac{2\pi}{\hbar} \sum_Q (n_Q + 1) |M_{K \to K'}|^2 \delta(|\delta E_{KK'}| - \hbar w_Q) \quad \text{(with phonon emssion)}.$$

Here, $n_Q$ = *average phonon occupation number*, and $M_{K' \to K}$ and $M_{K \to K'}$ are the effective matrix elements describing the 2$^{nd}$-order processes in Eqns. (VII-6) and (VII-7), respectively, with

$$|M_{K' \to K}| \approx |M_{K \to K'}| \approx \frac{1}{|\delta E_{KK'}|} |M_{QD}| \left| M_{EP}^{(K)}(Q) - M_{EP}^{(K')}(Q) \right|. \quad \text{(VII-9)}$$

For $B_{normal} = 0$, the EP part in Eqn. (VII-9) vanishes due to the time reversal symmetry. For a finite $B_{normal}$, the EP part is finite, and an upper bound estimate is given below. (A more accurate estimate is given later.) We write

$$|M_{EP}^{(K)}(Q) - M_{EP}^{(K')}(Q)| \leq O(|M_{EP}^{(K)}(Q)|) \quad [\text{or } O(|M_{EP}^{(K')}(Q)|)], \quad \text{(VII-10)}$$

$$|M_{EP}^{(K)}(Q)| \approx |M_{EP}^{(K')}(Q)| \approx \frac{D}{1 + L^2 Q^2} \sqrt{\frac{\hbar |Q|}{2\rho_a A c_s}} (\equiv M_{EP}(Q)).$$

Using Eqn. (VII-10), we obtain

$$\frac{1}{\tau_{K \leftrightarrow K'}} \leq O(1) \frac{2\pi}{\hbar} \left( \frac{1}{\delta E_{KK'}} \right)^2 \sum_Q (n_Q + \frac{1}{2} \pm \frac{1}{2}) M_{QD}^2 M_{EP}^2 \delta(|\delta E_{KK'}| - \hbar w_Q). \quad \text{(VII-11)}$$

After performing the Q-summation, we obtain

$$\frac{1}{\tau_{K\leftrightarrow K'}} \leq O(1) \frac{V_0^2 D^2}{\hbar^3 c_s^4 \rho_a \left(1+L^2 \delta k^2\right)^2 \left(1+L^2 Q_0^2\right)^2} (n_{Q_0} + \frac{1}{2} \pm \frac{1}{2}), \ \hbar c_s Q_0 = |\delta E_{KK'}|. \tag{VII-12}$$

Now, we give a more accurate estimate of the relaxation rate for weak $B_{normal}$. We expect the EP part, $[M_{EP}^{(K)}(Q) - M_{EP}^{(K')}(Q)]$, in Eqn. (VII-9) to be linear in $B_{normal}$ at weak $B_{normal}$. Therefore, for the estimate, we need to calculate each of the EP matrix elements correct to the order of $B_{normal}$. According to Eqn. (VII-3), this requires the calculation of the states $|K_{QD}\rangle$ and $|K_{QD}'\rangle$ correct to the same order.

We use the Hamiltonian given earlier and listed again here,

$$H \simeq H_{B\tau}^{(0)} + H_{B0}^{(0)} + H_{0\tau}^{(1)} + H_{B\tau}^{(1)} + H_{rest}, \tag{V-4}$$

$$H_{B\tau}^{(0)} = \frac{\vec{p}^2}{2m^*} + V + \tau_v \mu_{v0} B_{normal}, \ H_{B0}^{(0)} = \frac{e\vec{A}\cdot\vec{p}}{m^*},$$

$$H_{0\tau}^{(1)} = \tau_v \frac{\hbar}{4m^*\Delta}\left[(\nabla V)\times \vec{p}\right]_{normal}, \ H_{B\tau}^{(1)} = -\tau_v \mu_{v0}\left(B_{normal}\frac{\vec{p}^2}{2m^*\Delta} + \frac{1}{2\Delta}\left[\vec{A}\times(\nabla V)\right]_{normal}\right).$$

We take $V=V_{QD}(r) = \tfrac{1}{2}m^* w_0^2 r^2$. A perturbative calculation using the symmetric gauge $(A_x, A_y) = \tfrac{1}{2}(-B_{normal} y, B_{normal} x)$ yields the following ground state wave function [$\Phi_{00}(\tau_v=1)$ for the state $|K_{QD}\rangle$ and $\Phi_{00}(\tau_v=-1)$ for $|K_{QD}'\rangle$] correct to $O(B_{normal})$,

$$\phi_{00}(\tau_v) = \phi_{00}^{(0)} + \delta\phi_{B\tau} + \delta\phi_{rest}, \tag{VII-13}$$

$$\delta\phi_{B\tau} = \sum_{nm\neq 00}\frac{<\phi_{nm}^{(0)}|H_{B\tau}^{(1)}|\phi_{00}^{(0)}>}{E_{00}^{(0)}(\tau_v)-E_{nm}^{(0)}(\tau_v)}\phi_{nm}^{(0)} = -\frac{3\sqrt{2}}{16}\tau_v \frac{\mu_{v0}B_{normal}}{\Delta}\left(\phi_{20}^{(0)}+\phi_{02}^{(0)}\right),$$

$$H_{B\tau}^{(0)}\phi_{nm}^{(0)} = E_{nm}^{(0)}\phi_{nm}^{(0)}.$$

$\Phi_{nm}^{(0)}$ is the 2D isotropic oscillator wave function. $\delta\Phi_{rest}$ is the part of perturbative correction that does not depend on either $\tau_v$ or $B_{normal}$. Since $[M_{EP}^{(K)}(Q) - M_{EP}^{(K')}(Q)]$ is expected to be linear in $B_{normal}$, $\delta\Phi_{rest}$ is irrelevant for the calculation. Using $\Phi_{00}(\tau_v)$ in Eqn. (VII-13), we calculate $[M_{EP}^{(K)}(Q) - M_{EP}^{(K')}(Q)]$ according to Eqn. (VII-3),

$$M_{EP}^{(K)}(Q) - M_{EP}^{(K')}(Q) \approx \frac{3}{8} D \frac{\hbar}{m^* w_0} \frac{\mu_{v0} B_{normal}}{\Delta} |Q|^3 e^{-\frac{\hbar Q^2}{4m^* w_0}} \sqrt{\frac{\hbar}{2\rho_a A w_Q}} \ . \tag{VII-14}$$

When we substitute this result into Eqn. (VII-9), and (VII-9) into (VII-8), it gives

$$\frac{1}{\tau_{K\leftrightarrow K'}} \approx O(1)(n_{Q_0} + \frac{1}{2} \pm \frac{1}{2}) \frac{\mu_{v0}^2}{e^2 \hbar^5 \rho_a c_s^8} \frac{1}{(1+\delta k^2 L^2)^2} \frac{V_0^2 D^2}{(\hbar w_0 \Delta)^2} e^{-\frac{\hbar Q_0^2}{2m^* w_0}} (\mu_{v0} B_{normal})^6 \tag{VII-15}$$

("+" for phonon emission, "-" for phonon absorption)
$\hbar c_s Q_0 = |\delta E_{KK'}|$,
(for $B_{normal} < \frac{\hbar w_0}{2\mu_{v0}}$).

The condition $B_{normal} < \hbar w_0/2\mu_{v0}$ is imposed here to avoid level crossing between the lowest K-orbital and the 1st excited K'-orbital. For $\Delta = 0.14$eV, $L \sim 350$Å, $\hbar w_0 \sim 30$meV, this condition is satisfied for $B_{normal} < 6.7$T. Using these parameters, together with $V_0 \sim 0.5\Delta$ (for a bound state to exist), $D = 18$eV,[34] $c_s = 2.1 \times 10^4$m/s,[34] $B_{normal} = 100$mT, and temperature = 10K, we obtain $\tau_{K\leftrightarrow K'} \sim O$(ms), sufficiently long for qubit manipulation.

The discussion of valley relaxation can be generalized to

the singlet-triplet transition, $|z_S\rangle \leftrightarrow |z_{T_\pm}\rangle$, in the DQD structure

It yields

$$\frac{1}{\tau_{S\leftrightarrow T_\pm}} = \frac{2\pi}{\hbar} \sum_Q (n_Q + \frac{1}{2} \mp \frac{1}{2}) |M_{S\to T_\pm}|^2 \delta(|\delta E_{ST\pm}| - \hbar w_Q). \tag{VII-16}$$

Here, $\delta E_{ST\pm} = E_S - E_{T\pm}$ (singlet-triplet energy difference). The singlet and triplet states are listed earlier in **Sec. II**.

The transition, $|z_S\rangle \to |z_{T\pm}\rangle$, is also a 2$^{nd}$-order process (similar to those given in Eqns. (VII-6) and (VII-7)) and involves the phonon-mediated valley flip. $M_{S\to T+}$ and $M_{S\to T-}$ in Eqn. (VII-16) are the corresponding effective matrix elements of the process, with

$$M_{S\to T+} = \frac{1}{2\sqrt{2}} \frac{1}{\delta E_{ST+}} (M_{QD,R}^{(KK')} - M_{QD,L}^{(KK')}) \tag{VII-17}$$
$$\left[ M_{EP,L}^{(K)}(Q) + M_{EP,R}^{(K)}(Q) - M_{EP,L}^{(K')}(Q) - M_{EP,R}^{(K')}(Q) \right],$$

and

$$M_{S\to T-} = \frac{1}{2\sqrt{2}} \frac{1}{\delta E_{ST-}} (M_{QD,L}^{(K'K)} - M_{QD,R}^{(K'K)})$$
$$\left[ M_{EP,L}^{(K)}(Q) + M_{EP,R}^{(K)}(Q) - M_{EP,L}^{(K')}(Q) - M_{EP,R}^{(K')}(Q) \right].$$

The matrix elements appearing here are defined by expressions similar to those in Eqns. (VII-3) and (VII-4), modified to take into account that there are now left and right QDs involved, e.g.,

$$M_{EP,\chi}^{(K)}(Q) = D|Q| \sqrt{\frac{\hbar}{2\rho_a A w_Q}} \langle K_\chi | e^{i\vec{Q}\cdot\vec{r}} | K_\chi \rangle, \tag{VII-18}$$

$$M_{EP,\chi}^{(K')}(Q) = D|Q| \sqrt{\frac{\hbar}{2\rho_a A w_Q}} \langle K_\chi' | e^{i\vec{Q}\cdot\vec{r}} | K_\chi' \rangle,$$

$$M_{QD,\chi}^{(KK')} = \langle K_\chi | V_{QD}^{(\chi)} | K_\chi' \rangle, \quad M_{QD,\chi}^{(K'K)} = \langle K_\chi' | V_{QD}^{(\chi)} | K_\chi \rangle,$$

$\chi = L(left) \ or \ R(right) \ (QD).$

$V_{QD}^{(L)}$ and $V_{QD}^{(R)}$ are the confinement potentials of the left and right QDs, respectively.

If the left and right QDs are identical, the QD potential part in $M_{S\to T\pm}$ vanishes. In general, for dissimilar left and right QDs (with comparable size ~ L), we may write

$$\left|M_{S\to T\pm}\right| < O(1)\frac{1}{\left|\delta E_{ST\pm}\right|} M_{QD}\left|M_{EP}^{(K)}(Q) - M_{EP}^{(K')}(Q)\right|. \tag{VII-19}$$

This is analogous to the matrix element in Eqn. (VII-9) for valley flip, and, when used in Eqn. (VII-16), leads to Eqn. (VII-15) [with the substitution $\delta E_{KK'} \to \delta E_{ST\pm}$] as an upper bound estimate of $1/\tau_{S\leftrightarrow T\pm}$, giving $\tau_{S\leftrightarrow T\pm} > O(ms)$.

Similar calculation applies to

the transition $|z_S\rangle \leftrightarrow |z_{T\pm}\rangle$ in a single QD

This case is relevant, if the valley qubit is initialized by utilizing one-QD singlets, as in the case of spin-pair scheme.[13] The transition rate is again described by Eqn. (VII-16), and the matrix element involved is discussed in the following.

The singlet and triplet states are written below,

$$|z_S\rangle = c_{QD}^{(K_0)+} c_{QD}^{(K_0')+} |vaccum\rangle, \tag{VII-20}$$
$$|z_{T+}\rangle = c_{QD}^{(K_0)+} c_{QD}^{(K_1)+} |vaccum\rangle, \quad |z_{T-}\rangle = c_{QD}^{(K_1')+} c_{QD}^{(K_0')+} |vaccum\rangle.$$

Here, the index "0" represents the lowest orbital, and "1" the 1$^{st}$ excited one. The effective matrix element for the transition is

$$M_{S\to T+} = \frac{1}{\delta E_{ST+}} M_{QD}^{(K_1 K_0')}\left[M_{EP}^{(K_1)}(Q) - M_{EP}^{(K_0')}(Q)\right], \tag{VII-21}$$

$$M_{S\to T-} = \frac{1}{\delta E_{ST-}} M_{QD}^{(K_1' K_0)}\left[M_{EP}^{(K_1')}(Q) - M_{EP}^{(K_0)}(Q)\right]$$

$$M_{QD}^{(K_1 K_0')} = \langle K_1 | V_{QD} | K_0' \rangle, \quad M_{QD}^{(K_1' K_0)} = \langle K_1' | V_{QD} | K_0 \rangle$$

$$M_{EP}^{(K_i)}(Q) = DQ\sqrt{\frac{\hbar}{2\rho_a A w_Q}} <K_i | e^{i\vec{Q}\cdot\vec{r}} | K_i>,$$

$$M_{EP}^{(K_i')}(Q) = DQ\sqrt{\frac{\hbar}{2\rho_a A w_Q}} <K_i' | e^{i\vec{Q}\cdot\vec{r}} | K_i'>,$$

$i = 0, 1.$

For a symmetric QD in the absence of magnetic field, the lowest and the 1$^{st}$ excited orbitals have opposite parity symmetry. It follows that $M_{QD}^{(K1K0')} = M_{QD}^{(K1'K0)} = 0$ and, thereby, $M_{S\to T+}$ and $M_{S\to T-}$ both vanish. In the general case where the QD is asymmetric, we employ the following approximation

$$|M_{S\to T\pm}| \to O(1)\frac{1}{|\delta E_{ST\pm}|} M_{QD} M_{EP} \tag{VII-22}$$

This leads to Eqns. (VII-11) and (VII-12) [with the substitution $\delta E_{KK'} \to \delta E_{ST\pm}$] as an estimate for $1/\tau_{ST\pm}$. Using $\Delta \sim 0.14$eV, $L \sim 350$Å, $V_0 \sim 0.5\Delta$ (for a bound state to exist), $\delta E_{ST\pm} \sim 30$meV, $B_{normal} \sim 100$mT, and temperature = 10K, we obtain $\tau_{S\leftrightarrow T\pm} \gg O(ms)$.

Last, we discuss shortly

the singlet-triplet transition $|z_S\rangle \leftrightarrow |z_{T0}\rangle$ in the DQD

The transition is caused only by the EP interaction, and does not involve any valley flip. The energy difference between the states is J (the exchange splitting). The transition rate is given below,

$$\frac{1}{\tau_{S\leftrightarrow T_0}} = \frac{2\pi}{\hbar} \sum_Q (n_Q + \frac{1}{2} \pm \frac{1}{2}) |M_{S\to T_0}|^2 \delta(J - \hbar w_Q), \tag{VII-23}$$

("+" for $T_0 \to S$, "−" for $S \to T_0$),

with the matrix element

$$M_{S \to T0} = \frac{1}{2}\left[ M_{EP,L}^{(K)}(Q) - M_{EP,R}^{(K)}(Q) - M_{EP,L}^{(K')}(Q) + M_{EP,R}^{(K')}(Q) \right].  \tag{VII-24}$$

For identical left and right QDs, the matrix element vanishes. In the general case where the QDs are dissimilar (with comparable size ~ L), we may follow the calculation that leads to Eqn. (VII-14), and write

$$M_{S \to T0} \approx O(1) D \frac{(\mu_{vL} - \mu_{vR}) B_{normal}}{m^* w_0^2} |Q|^3 \, e^{-\frac{\hbar Q^2}{4 m^* w_0}} \sqrt{\frac{\hbar}{2\rho_a A w_Q}}. \tag{VII-25}$$

This gives

$$\frac{1}{\tau_{S \leftrightarrow T_0}} \approx O(1)(n_{Q_0} + \frac{1}{2} \pm \frac{1}{2}) \frac{\mu_{v0}^2}{e^2 \hbar^5 \rho_a c_s^8} \frac{D^2 J^6}{(\hbar w_0)^4} e^{-\frac{\hbar Q_0^2}{2 m^* w_0}} \left[(\mu_{vL} - \mu_{vR}) B_{normal}\right]^2, \tag{VII-26}$$

$$\hbar c_s Q_0 = J.$$

We use Eqn. (VII-26) to estimate $\tau_{S \leftrightarrow T0}$. We take $(\mu_{vL}-\mu_{vR})B_{normal}$ ~ 1µeV, L ~ 350Å, D ~ 18eV, $\hbar w_0$ ~ 30meV, and temperature ~ 10K. For the stage of AC mode manipulation when the rotation $\check{R}_z$ is acting on the qubit state, we take J ~ 0.1meV. On the other hand, for the stage when the rotation $\check{R}_x$ is acting on the qubit state, we take J ~ 0.1µeV. In either case, it give $\tau_{S \leftrightarrow T0}$ >> O(ms).

In conclusion, the discussion in this section shows that the qubit coherence time is O(ms) in a typical case.

## VIII. Summary

In summary, this work has examined the physics underlying valley pair qubits in graphene DQDs. Confined valley magnetic moments can be modulated electrically in DC or AC modes based on the analogue of 1st-order relativistic effect in graphene, with the distinctive advantage of being state-mixing free. This mechanism, together with the electrically tunable exchange coupling, allows all-electric manipulation of qubits via electric gates, on the time scale of ns. Moreover, the qubits envisioned here are fault-tolerant, with long coherence time ~ O(ms).

Implementation of valley pair qubits requires the capabilities to 1) open a gap in graphene, and 2) manufacture gated graphene devices in the nanometer scale, both being important issues in the current development of graphene-based nanoelectronics. Experimental realization of the qubits will lead to utilization of the valley d.o.f., in addition to the spin d.o.f., in quantum information encoding, and raise the interesting prospect of valley-based quantum computing / communication in carbon systems.

**Acknowledgment** – We thank the support of ROC National Science Council through the Contract No. NSC-99-2112-M-007-019.

## Appendix A

We sketch the derivation of the Schrodinger type equation in gapped graphene, with the magnetic effect and the 1$^{st}$-order R.C. included.

We begin with the Dirac-type equation

$$H_D \phi_D = E \phi_D, \tag{A-1}$$

$$H_D = \begin{pmatrix} \Delta + V & v_F \hat{\pi}_- \\ v_F \hat{\pi}_+ & -\Delta + V \end{pmatrix}, \phi_D = \begin{pmatrix} \phi_A \\ \phi_B \end{pmatrix},$$

$$\vec{\pi} = \vec{p} + e\vec{A}, \vec{A} = \frac{1}{2}(-B_{normal} y, B_{normal} x), \hat{p}_+ = \hat{p}_x + i\tau_v \hat{p}_y, \hat{p}_- = \hat{p}_x - i\tau_v \hat{p}_y,$$

$$A_+ = A_x + i\tau_v A_y, A_- = A_x - i\tau_v A_y, \hat{\pi}_+ = \hat{p}_+ + eA_+, \hat{\pi}_- = \hat{p}_- + eA_-.$$

Solving (A-1), we obtain the following relation between the two components of $\Phi_D$,

$$\phi_B = \frac{1}{\Delta + E - V} v_F \hat{\pi}_+ \phi_A, \tag{A-2}$$

and the following equation for the component $\Phi_A$,

$$H_N \phi_A = (E - \Delta) \phi_A, \tag{A-3}$$

$$H_N = v_F \hat{\pi}_- \left( \frac{1}{\Delta + E - V} v_F \hat{\pi}_+ \right) + V.$$

Eqn. (A-3) contains full relativistic effects. It can be reduced to the Schrodinger equation with only the 1$^{st}$-order R.C. included, as follows.

We expand $\frac{1}{\Delta + E - V}$ in (A-3), and keep the R.C. up to the 1$^{st}$ order,

$$H_N \phi_A \approx v_F \hat{\pi}_- \frac{1}{2\Delta} \left[ 1 - \frac{E - \Delta - V}{2\Delta} \right] v_F \hat{\pi}_+ \phi_A + V \phi_A,$$

$$= \left( H_N^{(0)} + H_N^{(1)} \right) \phi_A.$$

$$H_N^{(0)} \phi_A = v_F \hat{\pi}_- \frac{1}{2\Delta} v_F \hat{\pi}_+ \phi_A + V \phi_A \tag{A-4}$$

$$H_N^{(1)} \phi_A = -v_F \hat{\pi}_- \frac{1}{2\Delta}\left(\frac{E-\Delta-V}{2\Delta}\right) v_F \hat{\pi}_+ \phi_A \quad \text{(1st-order R.C.)} \tag{A-5}$$

$H_N^{(0)}$ is the non-relativistic part of the Hamiltonian, and $H_N^{(1)}$ is the 1$^{st}$-order R.C.. $H_N^{(0)}$ in Eqn. (A-4) is further simplified with the following identity,

$$\hat{\pi}_-(\hat{\pi}_+ f) = \hat{\pi}^2 f + 2m^* \tau_v \mu_{v0} B_{normal} f \quad (f = \text{generic expression}) \tag{A-6}$$

This gives

$$H_N^{(0)} = \frac{\vec{\pi}^2}{2m^*} + V + \tau_v \mu_{v0} B_{normal}, \tag{A-7}$$

$H_N^{(1)}$ in Eqn. (A-5) is also simplified, with the following identity,

$$(E-\Delta-V)\hat{\pi}_+ \phi_A = \hat{\pi}_+\left((E-\Delta-V)\phi_A\right) + (\hat{p}_+ V)\phi_A,$$

where the 1$^{st}$ term on the right hand side is further replaced, e.g., $(E-\Delta-V)\phi_A \to v_F \hat{\pi}_-\left(\frac{1}{\Delta+E-V} v_F \hat{\pi}_+ \phi_A\right) \approx \frac{1}{2m^*}\hat{\pi}_-\hat{\pi}_+\phi_A$, according to Eqn. (A-3). This yields

$$H_N^{(1)} = -\frac{1}{2\Delta}\left(\frac{\vec{\pi}^2}{2m^*} + \tau_v \mu_{v0} B_{normal}\right)^2 + \frac{\hbar \tau_v}{4m^*\Delta}\left[(\nabla V) \times \vec{\pi}\right]_{normal} \tag{A-8}$$

$$-\frac{e}{4m^*\Delta}\vec{A}\cdot(\vec{p}V) - \frac{1}{4m^*\Delta}(\vec{p}V)\cdot\vec{p} - \frac{1}{4m^*\Delta}(\vec{p}^2 V),$$

Note that $H_N^{(1)}$ includes non-Hermitian terms, such as $\frac{1}{4m^*\Delta}(\vec{p}V)\cdot\vec{p}$. A similar situation also arises in the derivation of Schrodinger equation (with the 1$^{st}$-order R.C. included) from the Dirac equation, in relativistic quantum mechanics, the reason being that $\Phi_A$ is not really the corresponding Schrodinger wave function. It is a component of the Dirac wave function $\Phi_D$ and, moreover, not even normalized to the 1$^{st}$-order R.C.. We follow the standard procedure in the textbook (for example, J. J. Sakurai, *Advanced Quantum Mechanics*, Addison-Wesley, 1967) to deal with the situation. We apply the following similarity transformation,

$$\phi_A \to \phi = \Omega\phi_A, \quad H_N \to H = \Omega H_N \Omega^{-1}, \quad \Omega = 1 + \frac{1}{8m^*\Delta}\hat{\pi}^2. \tag{A-9}$$

It can be verified that the new function, $\Phi$, is normalized to the 1$^{st}$-order R.C. The transformation in (A-9) converts $H_N$ into the following Hermitian Hamiltonian, H, with

$$\begin{aligned} &H\phi = E\phi, \\ &H = H^{(0)} + H^{(1)}, \\ &H^{(0)} = \frac{\vec{\pi}^2}{2m^*} + V + \tau_v \mu_{v0} B_{normal}, \\ &H^{(1)} = R_1^{(1)} + R_2^{(1)}, \\ &R_1^{(1)} = \frac{\hbar \tau_v}{4m^*\Delta}[(\nabla V) \times \vec{\pi}]_{normal}, \quad \text{(VOI)} \\ &R_2^{(1)} = -\frac{1}{2\Delta}\left(\frac{\vec{\pi}^2}{2m^*} + \tau_v \mu_{v0} B_{normal}\right)^2 - \frac{1}{8m^*\Delta}(\vec{p}^2 V). \end{aligned} \tag{A-10}$$

(The electron energy E here is defined with respect to the conduction band edge.)

**Appendix B**

An alternative treatment of the AC electric effect on $\mu_v$ is provided here employing a moving reference frame. We consider a QD in the presence of both a normal magnetic field ($\mathbf{B}_{normal}$) and an in-plane AC electric field ($\varepsilon_x^{(AC)} \sin(w_{AC}t)$ in the x-direction).

Firstly, we investigate the case where the confining potential is dominantly quadratic but asymmetric with an anharmonic $x^3$ term,

$$H\phi \approx (H^{(0)} + H^{(1)})\phi,$$

$$V = V_{QD}(x,y,t) + e\varepsilon_x^{(AC)} x \sin(w_{AC}t),$$
$$V_{QD}(x,y,t) = V_2(x,y,t) + V_3(x,y,t),$$
$$V_2(x,y,t) = \frac{m^* w_0^2}{2}(x^2 + y^2), V_3(x,y,t) = \frac{m^* w_0^2}{3} k_{3x} x^3.$$

Here, $H^{(0)}$ and $H^{(1)}$ are those given earlier, e.g., in **Appendix A**. $k_{3x}$ characterizes the strength of $V_3$. We assume weak $k_{3x}$ ($k_{3x} << (m^* w_0 / \hbar)^{1/2}$), in order to facilitate the perturbative argument below. We also assume the adiabatic condition, namely,

$$w_{AC} << w_0,$$

meaning that the electron displacement occurs on a time scale much slower than that of the orbital motion.

The potential minimum location, $(x_0(t),0)$ ($x_0(t) = x_\varepsilon^{(AC)} \sin(w_{AC}t)$, $x_\varepsilon^{(AC)} = -\frac{e\varepsilon_x^{(AC)}}{m^* w_0^2}$), varies in time, due to the AC field. Let $\varepsilon_x^{(AC)}$ be weak ($e\varepsilon_x^{(AC)} << (\hbar m^* w_0^3)^{1/2}$). We write

$$V \approx V_2\left[x-x_0(t), y, t\right] + V_3\left[x-x_0(t), y, t\right] + V_{2x}\left[x-x_0(t), y, t\right] \tag{B-1}$$
$$V_{2x}(x, y, t) = k_{3x} x_0(t) m^* w_0^2 x^2,$$

correct to the order of $x_0$. $V_{2x}$ modifies the quadratic part of confining potential in the x-direction, giving the new frequency parameter

$$w_0 \to w_x = w_0(1+\frac{\delta w_x}{w_0}), \quad \delta w_x = k_{3x} x_0(t) w_0. \tag{B-2}$$

Because of the adiabatic condition, $w_{AC} \ll w_0$, the problem here can be regarded as quasi-time independent, on the time scale of the orbital motion. We are particularly interested in the energy shift caused by the AC field. For the study of $\mu_v$-tuning, we focus on the part of the shift, denoted as $\delta E_{LiVCES}$, which is linear in both $\varepsilon_x^{(AC)}$ and $B_{normal}$, as well as valley-dependent, i.e.,

$$\delta E_{LiVCES} \propto \tau_v \varepsilon_x^{(AC)} B_{normal}.$$

(LiVCES means "linear valley-contrasting energy shift".)

We reformulate the present problem in a moving reference frame, with the following transformation,

$$x \to x' = x - x_0(t), \quad f(x, y; t) \to f'(x', y; t),$$

($f$ = generic expression). It leads to the equation (in the moving frame)

$$H'\psi'(x', y'; t) = i\hbar \partial_t \psi'(x', y'; t),$$
$$H' = H^{(0)}{}' + V_{2x}' + H^{(1)}{}' - v_{x0}\hat{p}_{x'},$$
$$v_{x0} \equiv \partial_t x_0 \propto \varepsilon_x^{(AC)}.$$

With the adiabatic condition, we regard H' as being quasi-time independent, and discuss the effects of $\varepsilon_x^{(AC)}$-dependent terms, i.e., $V_{2x}'$ and $v_{x0}\hat{p}_{x'}$, within the time-independent perturbation theory. We write the corresponding time-independent wave equation

$$H'\phi'(x',y') = E\phi'(x',y'),$$
$$H' = H^{(0)}{}' + V_{2x}' + H^{(1)}{}' - v_{x0}\hat{p}_{x'}.$$

Firstly, we estimate the effect of the term, $v_{x0}\hat{p}_{x'}$. We note that

$$E(\tau_v, x_0, v_{x0}) = E(\tau_v, x_0, -v_{x0}).$$

[We regard $x_0$ and $v_{x0}$ as independent parameters in H'. Given $V'(x',y) = V'(x',-y)$ here, it can be verified that if $\Phi'(x',y)$ is a eigenstate of $H'(\tau_v, x_0, v_{x0})$ with energy E, then $\Phi'^*(x',-y)$ is an eigenstate of $H'(\tau_v, x_0, -v_{x0})$ with the same energy E.] It shows that the effect of the term on the energy is of the order of $v_{x0}^2$ ($\alpha\, \varepsilon_x^{(AC)\,2}$) or of a higher order, and does not contribute to $\delta E_{LiVCES}$. We thereby drop this term below. (From here on, we also drop the prime notation whenever it does not cause confusion.)

Thereby, we write

$$H \approx H^{(0)} + H^{(1)} + V_{2x}. \tag{B-3}$$

The presence of $V_{2x}$ shows that the quadratic part of $V_{QD}$ is modified, as described in Eqns. (B-1) and (B-2), into one with a slight anisotropy describing a rectangular QD. Now, Eqn. (V-5) for a rectangular QD can be applied. We obtain

$$\delta E_{LiVCES} = \tau_v \delta\mu_v^{(AC)} B_{normal}, \tag{B-4}$$
$$\delta\mu_v^{(AC)} = -\frac{1}{8}\mu_{v0}\left(k_{3x}x_\varepsilon^{(AC)}\right)\sin(w_{AC}t)\left(\frac{\hbar w_0}{\Delta}\right), \quad \text{(cubic case)}$$

in agreement with Eqn. (VI-7). Note that the presence of the additional $x^3$ term (in $H^{(0)}$) in Eqn. (B-3) does not affect our leading-order estimate of $\delta E_{LiVCES}$. It can be shown that this term enters the perturbative correction at a higher order.

We extend the result here to the quadric case where

$$V_{QD} = V_2 + V_4,$$

$$V_2 = \frac{1}{2} m^* w_0^2 (x^2 + y^2), V_4(x) = \frac{1}{4} k_{4x} m^* w_0^2 x^4.$$

Asymmetry in the potential can be introduced by applying a DC gate voltage,

$$V = V_{QD} + e\varepsilon_x^{(DC)} x,$$

$$\approx \frac{1}{2} m^* w_0^2 \left[ (x - x_\varepsilon^{(DC)})^2 + y^2 \right] + \frac{1}{4} k_{4x} m^* w_0^2 (x - x_\varepsilon^{(DC)})^4 + k_{4x} m^* w_0^2 x_\varepsilon^{(DC)} (x - x_\varepsilon^{(DC)})^3,$$

$$x - x_\varepsilon^{(DC)} \to x$$
$$\to$$

$$V = \frac{1}{2} m^* w_0^2 \left( x^2 + y^2 \right) + \frac{1}{3} k_{3x} m^* w_0^2 x^3 + \frac{1}{4} k_{4x} m^* w_0^2 x^4, \qquad (B-5)$$

$$k_{3x} = 3 k_{4x} x_\varepsilon^{(DC)}, \quad x_\varepsilon^{(DC)} = -\frac{e\varepsilon_x^{(DC)}}{m^* w_0^2}.$$

Here we assume $\varepsilon_x^{(DC)}$ is weak ($e\varepsilon_x^{(DC)} \ll (\hbar m^* w_0^3)^{1/2}$ but $\varepsilon_x^{(DC)} \gg \varepsilon_x^{(AC)}$), and retain only the terms up to the order of $x_\varepsilon^{(DC)}$. Eqn. (B-5) shows that the DC field produces a cubic term with the strength $k_{3x} = 3 k_{4x} x_\varepsilon^{(DC)}$. Using the expression of $\delta E_{LiVCES}$ derived earlier in the cubic case, we obtain

$$\delta E_{LiVCES} = \tau_v \delta\mu_v^{(AC)} B_{normal}, \qquad (B-6)$$

$$\delta\mu_v^{(AC)} = -\frac{3}{8} \mu_{v0} \left( k_{4x} x_\varepsilon^{(DC)} x_\varepsilon^{(AC)} \right) \sin(w_{AC} t) \left( \frac{\hbar w_0}{\Delta} \right), \quad \text{(quadric case)}$$

in agreement with Eqn. (VI-8). Note that the inclusion of the $x^4$ term in the potential in Eqn. (B-5) does not affect our leading-order estimate of $\delta E_{LiVCES}$. It can be shown that this term enters the perturbative correction at a higher order.

**Figure Captions**

**Figure 1.** The DQD qubit structure. The QDs are electrostatically defined, for example, by back gates (not shown in the figure). Gate $V_C$ is used to tune the potential barrier. Gates $V_L$ and $V_R$ are applied for DC/AC $\mu_v$-tuning.

**Figure 2.** One-electron energy levels in a QD placed in the tilted magnetic field $\mathbf{B}_{total}$ ($\mathbf{B}_{total} = \mathbf{B}_{plane} + \mathbf{B}_{normal}$). Conduction and valence bands of gapped graphene are also shown. The Zeeman-type interaction for spin and valley d.o.f.s, $H_Z = -g^* \sigma \mu_B |\mathbf{B}_{total}| + \tau_v \mu_v |\mathbf{B}_{normal}|$, splits the quadplet $|\tau_v = \pm 1, \sigma = \pm 1/2\rangle$. For $g^* \mu_B |\sigma \mathbf{B}_{total}| > \mu_v |\mathbf{B}_{normal}|$, the splitting leaves $|\tau_v = \pm 1, \sigma = 1/2\rangle$ as the lower valley doublet.

**Figure 3.** Two-electron energy levels in the DQD. $|z_{T\pm}\rangle$ are split away from $\{|z_S\rangle, |z_{T0}\rangle\}$, by $\pm(\mu_{vL} + \mu_{vR})|\mathbf{B}_{normal}|$, respectively. $|z_S\rangle$ and $|z_{T0}\rangle$ are split in energy by J, while $|x_-\rangle$ and $|x_+\rangle$ by $2(\mu_{vL} - \mu_{vR})|\mathbf{B}_{normal}|$.

**Figure 4.** The time evolution of a qubit state, as governed by $H_{eff}$, consists of a rotation $\check{R}_x(\theta_x)$ about the x-axis of the Bloch sphere, and a rotation $\check{R}_z(\theta_z)$ about the z-axis. $\theta_x$ and $\theta_z$ are the respective angles of rotation.

**Figure 5.** In the AC mode, the initial qubit state, e.g., $|z_S\rangle$, may be manipulated in the alternating sequence, $\check{R}_x(\theta_x^{(AC)}) \rightarrow \check{R}_z(\theta_z=\pi) \rightarrow \check{R}_x(-\theta_x^{(AC)}) \rightarrow \check{R}_z(\theta_z=\pi) \rightarrow ..... \check{R}_z(\theta_z^{(target)}+\pi/2)$, into a target state ($\theta_z^{(target)} =$ *target state longitude*).

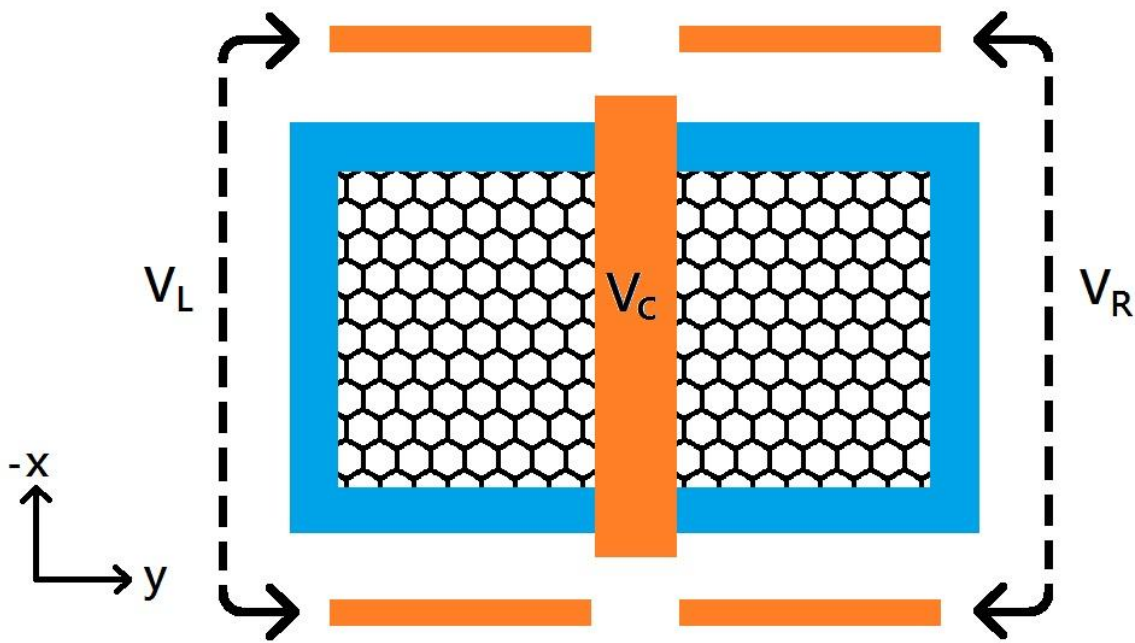

Figure 1

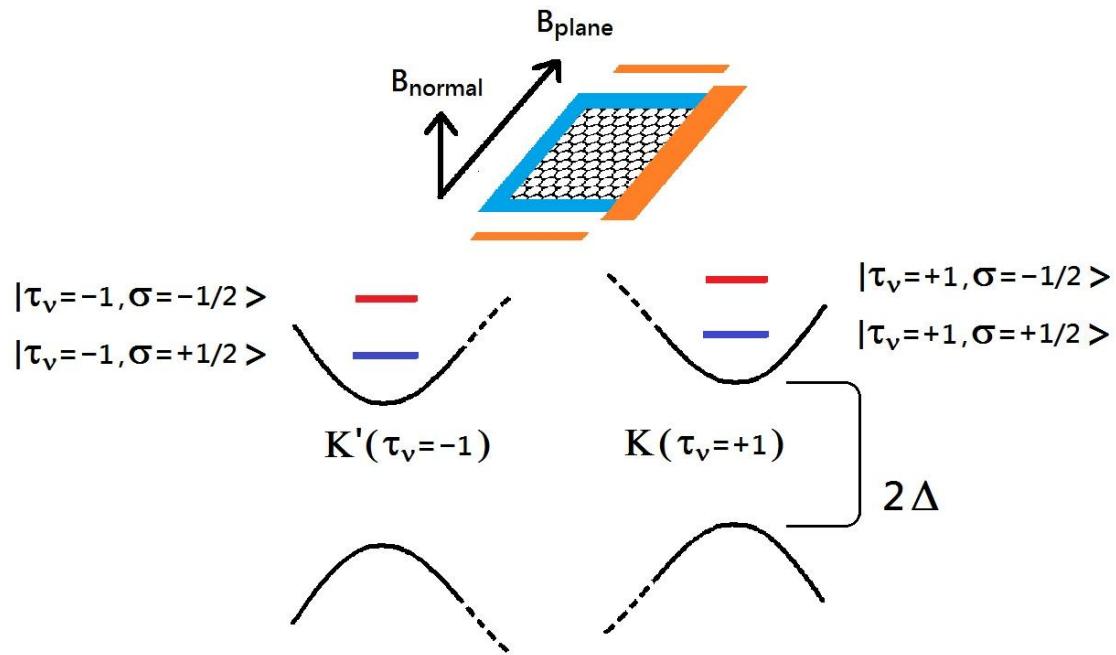

Figure 2

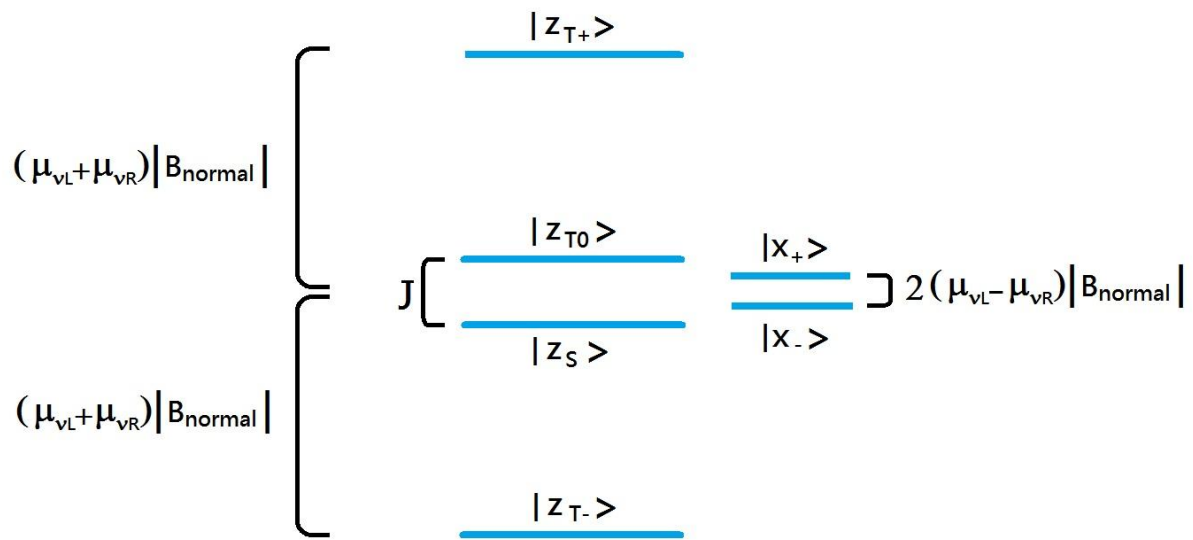

Figure 3

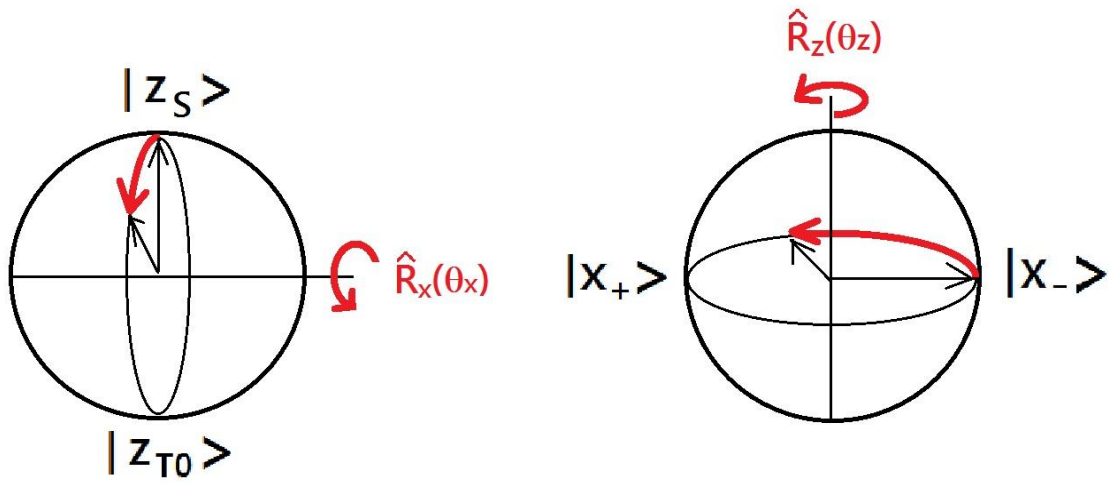

Figure 4

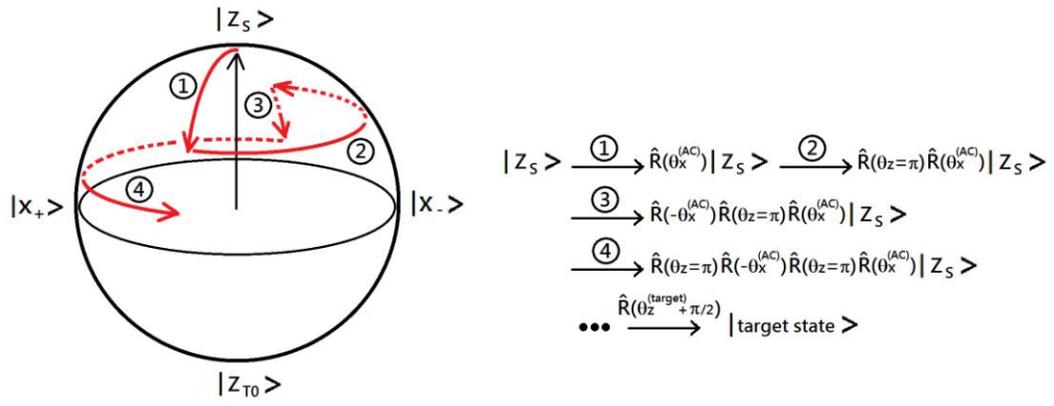

Figure 5